\documentstyle[preprint,aps]{revtex}
\begin{document}
\draft
\title{Quantum Evaporation from Superfluid Helium at Normal Incidence}
\author{F. Dalfovo$^1$, M. Guilleumas$^1$,
A. Lastri$^1$, L. Pitaevskii$^2$ and S. Stringari$^1$}
\address{$^1$Dipartimento di Fisica, Universit\`a di Trento,
I-38050 Povo, Italy}
\address{$^2$Department of Physics, TECHNION, Haifa 32000, Israel,
and \\
\mbox{
Kapitza Institute for Physical Problems, ul. Kosygina 2, 117334 Moscow}}
\maketitle
\begin{abstract}
We study the scattering of atoms, rotons and phonons at the free surface
of $^4$He at normal incidence and calculate the evaporation, condensation
and reflection probabilities.
Assuming elastic one-to-one processes and using general properties of
the scattering matrix, such as unitarity  and time reversal, we argue
that all nonzero probabilities can be written in terms of a single
energy-dependent parameter.
Quantitative predictions are obtained using
linearized time dependent density functional theory.
\end{abstract}

Superfluid helium at low temperature exhibits the peculiar phenomenon
of quantum evaporation. Elementary excitations, such as rotons and high
energy phonons, propagate in the liquid with long mean free path and
with an energy which can exceed the binding energy of atoms in the liquid.
Thus, when an elementary excitation impinges upon the free surface it may
eject an atom out of the liquid through a quantum process. The scattering
of excitations and atoms at the free surface of the superfluid
have been extensively studied in the recent years (see for instance
Refs.\cite{Brown,Tuck,QC} and references therein), but  the comparison
between theory and experiments is still unsatisfactory. Previous
theoretical approaches to the problem of quantum evaporation made use
of semiclassical approximations  \cite{Mar92,Mul92}. Recently 
\cite{PRL,JLTP} we have calculated the probability of evaporation,
condensation and reflection of rotons and atoms using a time dependent 
density functional (TDDF) theory. The density functional has proven a 
powerful method to investigate various structural features of inhomogeneous 
superfluid helium \cite{functional} including static and dynamic properties
of the free surface, droplets and films. When applied to quantum 
evaporation, the linearized TDDF theory accounts only for 
one-to-one processes, 
but includes  quantum effects beyond the semiclassical 
approximation.  So far we have restricted the analysis to the 
phonon forbidden region \cite{PRL,JLTP}, corresponding to
incidence angles such as phonons are excluded
by energy and momentum conservation. In the present work we apply
the same theory to the case  of scattering at normal incidence, where
atoms, rotons as well as phonons take part in the scattering processes.
We show that the
complexity of the scattering processes is dramatically reduced on the
basis of simple arguments based on symmetry properties of the scattering
matrix and threshold  effects.
The final result is that all nonzero probabilities can be expressed
in terms of a single energy-dependent
parameter. The numerical solution of the time
dependent density functional equations confirms this picture and provides
quantitative predictions for the probabilities as a function of energy.

In Fig.~1 we show the phonon-roton dispersion curve in bulk liquid.
The phonon branch goes from $q=0$ up to the maxon ($\Delta^*$).
On the right of the maxon the slope is negative
and corresponds to the dispersion of $R^-$ rotons, that is, rotons
with negative group velocity. Then the slope becomes positive again
for $R^+$ rotons on the right of the roton minimum ($\Delta$).
The threshold for atom evaporation at normal incidence coincides with
the chemical potential, $|\mu|=7.15$ K (dashed line):
only excitations with energy
higher than $|\mu|$ can produce quantum evaporation. The energy of an
evaporated atom is simply $|\mu| + \hbar^2 q^2 /2m$.
For scattering at normal incidence, the problem is unidimensional;
the wave vector ${\bf q}$ is orthogonal to the free surface and is not a
conserved quantity, since the surface breaks translational invariance.
As already said, we consider elastic processes in which the energy
$\hbar \omega$ is conserved.

Four types of excitations interact at the free surface of the superfluid:
atoms $(a)$, phonons $(p)$, rotons with positive $(+)$ and negative $(-)$
group velocity. One can define the scattering matrix by means of
$\Psi_{out}=S \Psi_{in}$, where $\Psi_{in}$ and $\Psi_{out}$ are the
incoming and outgoing asymptotic solutions, respectively.
The matrix element $S_{ij}$ connects a particular input ($i$) and
output ($j$) channel, with $i,j=a,p,+,-$, and depends 
on energy and incident angle.
It determines the probability associated with each scattering process
by $P_{ij}=|S_{ij}|^2 \,.$
These channels  correspond to atoms, phonons and rotons excitations states.
Thus, there are 16 complex scattering  matrix elements $S_{ij}$, and hence 
16  probabilities $P_{ij}$,  to be
determined. The unitarity and the time reversal symmetry ($t \to -t$)
of $S$ imply $S^\dagger = S^{-1}$  and $S^*= S^{-1}$,
respectively.  Combining these two conditions one finds that the scattering
matrix elements must satisfy the general property $S_{ij}=S_{ji}$,
reducing the independent matrix elements to 10.  Furthermore, the unitarity
condition $S^*S=1$ gives ten additional constraints. 
For instance, if the incident excitation
is of type $i$, then one has the unitarity conditions
\begin{equation}
\sum_{j} |S_{ij}|^2=1 \, ,
\label{sumj}
\end{equation}
where $i,j=a,p,+,-$, yielding four linear
combinations of matrix elements. The other six conditions are
\begin{equation}
S^*_{aa} S_{ap} + S^*_{ap} S_{pp} + S^*_{a-} S_{-p} + S^*_{a+} S_{+p} = 0
\label{unitarity*1}
\end{equation}
\begin{equation}
S^*_{aa} S_{a-} + S^*_{ap} S_{p-} + S^*_{a-} S_{--} + S^*_{a+} S_{+-} = 0
\label{unitarity*2}
\end{equation}
\begin{equation}
S^*_{aa} S_{a+} + S^*_{ap} S_{p+} + S^*_{a-} S_{-+} + S^*_{a+} S_{++} = 0
\label{unitarity*3}
\end{equation}
\begin{equation}
S^*_{pa} S_{a-} + S^*_{pp} S_{p-} + S^*_{p-} S_{--} + S^*_{p+} S_{+-} = 0
\label{unitarity*4}
\end{equation}
\begin{equation}
S^*_{pa} S_{a+} + S^*_{pp} S_{p+} + S^*_{p-} S_{-+} + S^*_{p+} S_{++} = 0
\label{unitarity*5}
\end{equation}
\begin{equation}
S^*_{-a} S_{a+} + S^*_{-p} S_{p+} + S^*_{--} S_{-+} + S^*_{-+} S_{++} = 0
\; .
\label{unitarity*6}
\end{equation}
We will now combine these rigorous relations with further arguments about
threshold effects and momentum exchange at the surface in order to reduce
the number of relevant unknowns.

The roton minimum  represents the threshold energy to excite
$R^-$ and $R^+$ rotons in the liquid. These excitations
coincide at $\Delta$ and, on the basis of symmetry arguments, one can
prove \cite{ripplons} that the mode-change reflection between $R^{-}$
and $R^+$ is the dominant one just above $\Delta$ ($P_{+-} \simeq 1$) and
that the other probabilities involving $R^{-}$ and $R^+$ are small. 
Conversely, below it only phonons and atoms are present. Among the
different processes the phonon $\leftrightarrow$ atom
scattering is expected to be favoured ($P_{pa}\simeq 1$), since it
implies the smallest change of momentum.  In fact, the normal reflection
of a  phonon or an atom would imply a much larger momentum transferred
to the surface and this seems unlikely for a smooth surface like the
one of helium, except near the threshold $\hbar \omega
\simeq |\mu|$
where the atom momentum goes to zero and full reflection takes place.
Now,  assuming  $P_{+-} \simeq 1$
and $P_{pa}\simeq 1$ into the unitarity conditions (\ref{sumj}), one
finds that all the remaining probabilities should vanish close
to $\Delta$.

Similarly, the maxon is a  threshold for phonons and  $R^-$ rotons and
one expects  the mode-change reflection between  $R^{-}$ and phonons
to dominate just below $\Delta^*$ ($P_{p-} \simeq 1$). Just above it,
one has only $R^+$ and atoms. For the same argument of smallest
momentum transferred to the surface, the $R^+ \leftrightarrow$ atom
scattering
should be much favoured ($P_{+a}\simeq 1$) with respect to the normal
reflection of  rotons and atoms. Putting $P_{p-} \simeq 1$ and
$P_{+a}\simeq 1$ in the unitarity conditions one finds again that
the other probabilities have to vanish near $\Delta^*$.

The above arguments suggest that $P_{pa}$ and $P_{+-}$ decrease from
$1$ to $0$ by increasing the energy from $\Delta$ to $\Delta^*$, while
$P_{p-}$ and $P_{+a}$ increase form $0$ to $1$ in the same
range. All probabilities should be smooth functions of the energy. The
ones that are $0$ both at $\Delta$ and $\Delta^*$ are expected to be
small everywhere in between. If we make the assumption that they
vanish for any value of energy ($\Delta\leq \hbar\omega\leq \Delta^*$),
\begin{equation}
P_{aa} = P_{pp} = P_{--} = P_{++} = P_{-a} = P_{+p} = 0 \; ,
\label{zero}
\end{equation}
we find simple and useful relations among the remaining nonzero
probabilities. In fact, by inserting assumption (\ref{zero}) in the
unitarity conditions (\ref{sumj}-\ref{unitarity*6}),
after some simple algebra one gets \cite{LTunitarity}
\begin{equation}
P_{pa}=P_{+-}=1-P_{p-}=1-P_{+a}  \; .
\label{P}
\end{equation}
This means that all nonzero probabilities at normal incidence can be 
written in terms of a single parameter. This result is expected to hold
for any theory accounting only for one-to-one elastic processes. 

In the remaining part of the paper we show that the numerical calculation
of $P_{ij}$, within time dependent density functional theory, are consistent
with the general properties of the scattering matrix and with
the threshold effect near $\Delta$ and $\Delta^*$. Furthermore, it
confirms the validity of assumption (\ref{zero}).

In order to calculate explicitly the probabilities $P_{ij}$ one needs
a theory for the elementary excitations of the nonuniform liquid.
The theory must provide a  spectrum of elementary excitations
close to the  experimental one, since  the proper energy balance
between the excitations is crucial;  moreover, it
must allow one to calculate  the asymptotic flux of elementary excitations
in a given process, in  order to extract the corresponding matrix elements.

In a density functional approach \cite{PRL,JLTP} one assumes the
energy of the system to be in the form
$E= \int d{\bf r} \,{\cal H}[\Psi,\Psi^*]$, where the complex function
$\Psi=\Phi \exp(i S/\hbar)$ is related to the density and velocity of the
superfluid by means of $\rho=\Phi^2$ and ${\bf v}=(1/m) {\bf \nabla} S$.
We use a  phenomenological energy density ${\cal H}$ \cite{functional}
which has been adjusted to provide an accurate description of the
equation of state, the static response function and the phonon-roton
dispersion law of the bulk liquid.  The description of the functional
can be found in our previous papers \cite{JLTP,functional}.

The eigenenergies and eigenfunctions of the system are calculated within
a time dependent density functional approach.
The propagation of the elementary excitations is described by linearizing
the wave function around the ground state
$\Psi_0({\bf r})$:
\begin{equation}
\Psi({\bf r},t)=\Psi_0({\bf r}) + \delta \Psi({\bf r},t)\, .
\label{dPsi}
\end{equation}
Far away from the surface $\delta \Psi({\bf r},t)$ corresponds to the
propagation of phonons and rotons in the liquid and free atoms in the
vacuum.  Due to linearization, only one-to-one
processes are taken into account. Thus the theory can not describe inelastic
processes, such as decay into multi-ripplons or multi-phonons.

For practical reasons we work in a slab geometry. The semi-infinite
system is simulated by slabs of liquid with thickness ranging from $50$ to
$100$ \AA, and centred in a finite box whose size is of the order
of $100 \div 150$ \AA.  The density functional  provides also
the ground state density profile of the liquid as the stationary solution
of lowest energy. The profile of the liquid-vacuum interface has a thickness
of about $6$ \AA\ and a shape close to the one obtained with
{\it ab initio} Monte  Carlo calculations (see Ref.~\cite{functional} for 
details). 
Choosing $z$ along the normal to the surface, one can write
the wave function $\delta \Psi$
in the form
\begin{equation}
\delta \Psi(z,t) = f(z) e^{-i\omega t} +
                         g(z) e^{i\omega t}\, .
\label{eq6}
\end{equation}
The quantities $f(z)$ and $g(z)$ are real functions which have to be
determined, together with $\omega$, by solving self-consistently
the equations of motion
\begin{equation}
\delta \int \! dt \! \int \! d{\bf r}\, \left\{
    {\cal H}[\Psi,\Psi^*] -
    \mu \Psi \Psi^* -
    \Psi^* i \hbar \frac{\partial \Psi}{\partial t} \right\}  = 0\,,
\label{eq2}
\end{equation}
linearized with respect to $f$ and $g$. Equations (\ref{eq6}) and (\ref{eq2})
assume the typical form of the random phase approximation
(RPA) for bosons.  In the formalism of RPA, $f(z)$ and $g(z)$ take
into account the particle-hole
and hole-particle transitions, respectively.  The two components
$f$ and $g$ of the excited states are coupled by
the equations of motion (\ref{eq2}).

We have solved numerically the linearized TDDF equations
in the finite box, obtaining  a set of discrete eigenenergies
$\omega$ and the corresponding eigenfunctions $f(z)$ and $g(z)$.
For a given $\omega$, the solution is a stationary state. Inside
the slab,  $f(z)$ and $g(z)$ are oscillating functions associated
with the propagation of phonons and rotons, whose dispersion 
law is shown in Fig.~1. 
Outside the slab, where particles are uncorrelated, the function
$g(z)$ vanishes while the equation for $f(z)$ coincides  with the
Schr\"odinger equation for free atoms. By looking at the
Fourier transforms of the signal far away from the surface,
both inside and outside the slab,  one can evaluate the asymptotic
amplitudes ($f_i$ and $g_i$) and the momentum $q_{i}$  of each type
of excitation which contributes to $\delta \Psi$.
A fitting procedure has been used in order to extract the
numerical values of $f_i$ and $g_i$ needed for the analysis of the
evaporation rates.

A given scattering process at a certain energy can be obtained as
a linear combination of different stationary solutions at that energy.
The latter can be found by slightly
varying the slab thickness ($L_{slab}$) and the box size ($L_{box}$).
Then, one can  evaluate the flux of incoming and outgoing  excitations.
In the present linearized TDDF theory, the current associated with a
given elementary excitation is
\begin{equation}
{\bf j}_i = {\bf v}_i (|f_i|^2-|g_i|^2) \; ,
\label{current}
\end{equation}
where ${\bf v}$ is its group velocity. From the
asymptotic fluxes one can easily evaluate the evaporation,
condensation and reflection probabilities \cite{JLTP}.

We have first verified that the resulting probabilities
$P_{ij}$  satisfy,  within the accuracy of the calculation,
the symmetry property  $P_{ij}=P_{ji}$ and the unitarity 
conditions (\ref{sumj}) in the whole range of energy 
$\hbar \omega > |\mu|$. Explicit results are shown in Table 
\ref{tab:combinations} for an intermediate value of energy 
($\hbar\omega=11$ K). Four states at the same energy are 
combined to describe the scattering processes
(atom $\to$ atom, phonon, $R^-$, $R^+$),
(phonon $\to$ atom, phonon, $R^-$,$R^+$),
($R^+$\ $\to$ atom, phonon, $R^-$,$R^+$) and
($R^-$ \ $\to$ atom, phonon, $R^-$,$R^+$).
The numerical uncertainty, which is less than $\pm 0.05$ for all 
probabilities, originate mainly from the fact that the states 
in the  linear combinations may be not enough linearly 
independent \cite{JLTP}. 

We find also that the probabilities $P_{aa}, P_{pp}, P_{--}, P_{++}, P_{-a}$
and $P_{p+}$ are extremely small at all energies; they turn out to be zero
within the present accuracy, confirming the hypothesis (\ref{zero}) made
before. Furthermore, the nonzero probabilities turn out to verify
relations (\ref{P}). It is therefore possible to resume all
the information about the nonzero probabilities in a single function
of energy, $P(\omega)$. Let us call $P(\omega) \equiv P_{+a}$.
This quantity is  plotted in Fig.~2, in the energy range
$\Delta \le \hbar \omega \le \Delta^*$.
The probability of mode-change scattering between
$R^-$ and phonons ($P_{p-}$) is equal to the evaporation probability
for $R^+$ rotons ($P_{+a}$); both probabilities start from $0$ at
the roton minimum and increase to $1$ at the maxon energy.
The evaporation probability for $R^+$ rotons is equal to $1$ even above
the maxon energy. The evaporation probability for phonons ($P_{pa}$) is
equal to the one for the mode-change process between rotons
($P_{+-}$). They are equal to $(1-P_{+a})$, so that they decrease
from $1$ to $0$ between $\Delta$ and $\Delta^*$.
It is worth noting that the
present results for the evaporation probability $P_{+a}$ at normal
incidence coincide, within the error bar, with the ones obtained previously
in the phonon forbidden region \cite{JLTP}, that is, for
$R^+$ rotons impinging
at about $15 \div 20$ degrees. The same is true for the mode-change
probability $P_{+-}$ \cite{LT21qe}.

Only few experimental estimates are available for the probabilities
$P_{ij}$. Indeed, by measuring time-of-flight and angular distribution
of evaporated atoms,  one-to-one evaporation processes have
been clearly seen \cite{Brown}, but a quantitative determination of
the ratio between incoming and outgoing fluxes, and hence $P_{ij}$, is
difficult. There are evidences for a sizable probability $P_{+a}$, which
should increase with a trend similar to the one in Fig.~1 \cite{Forbes}.
A recent estimate \cite{TuckLT} of the phonon $\rightarrow$ atom
probability, for high  energy phonons ($\hbar \omega > 10$K), is
$P_{pa}\simeq 0.1$. Our theory, in the same range of energy, gives
a value which decreases rapidly from $0.5$ to $0$. We also find that an
incident atom condenses with probability $P_{ap}+P_{a+}\simeq 1$ and
hence the reflection probability $P_{aa}$ is almost zero. This is
in agreement with the observed small reflectivity \cite{Tuck,QC}; but the
experiments support also the idea that the atom condensation might
occur  through processes like two-phonon or multi ripplon
production.  The extension of the
present formalism to include such mechanisms, beyond the one-to-one
hypothesis, remains an important task, for a more systematic
comparison between theory and experiments.

\vskip 1 cm

M. G. thanks the Generalitat de Catalunya for financial support.

\begin{figure}
\caption{
Phonon-roton dispersion in liquid $^4$He. The dashed line is the
threshold $|\mu|=7.15$ K for atom evaporation.
}
\label{fig1}
\end{figure}
\begin{figure}
\caption{
Probability $P \equiv P_{+a}$ as a function of energy.
The other nonzero
probabilities can be extracted using Eq~(\protect\ref{P}).
Energy scale starts from $\Delta$.
}
\label{fig2}
\end{figure}

\begin{table}
\caption{Probabilities $P_{ij}$ for four different scattering processes
(incident atom, phonon, $R^-$ and $R^+$, respectively) described by linear
combinations of four states at the same energy ($11$ K) and 
for different values of ($L_{slab}$,$L_{box}$). The unitarity 
condition, Eq.~(\protect\ref{sumj}), is labeled by $\Sigma$.}
\begin{tabular}{c|cccc|c}
\multicolumn{1}{c|}{incident exc.}&
\multicolumn{4}{c|}{$P_{ij}$}&
\multicolumn{1}{c}{$\Sigma$}\\
\hline
atom & $P_{aa}$=0.0002 & $P_{ap}$=0.3384 & $P_{a-}$=0.0003 &
$P_{a+}$=0.6837 & 1.0226 \\ \hline
phonon & $P_{pa}$=0.3236 & $P_{pp}$=0.0020 & $P_{p-}$=0.6487 &
$P_{p+}$=0.00002 & 0.9743 \\ \hline
$R^-$ & $P_{-a}$=0.0004 & $P_{-p}$=0.6859 & $P_{--}$=0.0013 &
$P_{-+}$=0.3408 & 1.0284 \\ \hline
$R^+$ & $P_{+a}$=0.6537 & $P_{+p}$=0.0001 & $P_{+-}$=0.3222 &
$P_{++}$=0.00002 & 0.9760 \\
\end{tabular}
\label{tab:combinations}
\end{table}


\begin{thebibliography}{99}

\bibitem{Brown}
Brown M and Wyatt A F G 1990 J. Phys.: Condens. Matter {\bf 2} 5025

\bibitem{Tuck}
Wyatt A F G, Tucker M A H  and Cregan R F 1995 Phys. Rev. Lett. {\bf 74}
5236

\bibitem{QC}
Edwards D O 1982 Physica {\bf 109} $\&$ {\bf 110B} 1531, and references
therein; Nayak V V, Edwards D O and Masuhara N 1983 Phys. Rev. Lett. {\bf 50}
990. Mukherjee S, Candela D, Edwards D O and
Kumar S 1987 Jap. J. Appl. Phys. {\bf 26-3} 257

\bibitem{Mar92} H. J. Maris, J. Low Temp. Phys. {\bf 87}, 773 (1992)

\bibitem{Mul92} P.A. Mulheran and J.C. Inkson, Phys. Rev. B {\bf 46},
5454 (1992)

\bibitem{PRL}
Dalfovo F, Fracchetti A, Lastri A, Pitaevskii L and Stringari S 1995
Phys. Rev. Lett. 75 2510

\bibitem{JLTP}
Dalfovo F, Fracchetti A, Lastri A, Pitaevskii L and Stringari S 1996
J. Low Temp. Phys. {\bf 104} 367

\bibitem{functional}
Dalfovo F, Lastri A, Pricaupenko L, Stringari S and Treiner J 1995
Phys. Rev. B {\bf 52} 1193

\bibitem{ripplons}
Lastri A, Dalfovo F, Pitaevskii L and Stringari S 1995
J. Low Temp. Phys. {\bf 98} 227

\bibitem{LTunitarity}
For a similar prodedure applied in the phonon forbidden region see
ref.~\protect\cite{JLTP} or:
Dalfovo F, Pitaevskii L and Stringari S 1996 {\it Proceedings of the
LT21 Conference},  Czech. J. Phys. {\bf 46-S1} 391

\bibitem{LT21qe}
Stringari S, Dalfovo F, Guilleumas M, Lastri A and Pitaevskii L
{\it Proceedings of the LT21 Conference},  Czech. J. Phys., in press

\bibitem{Forbes}
Forbes A C and Wyatt A F G 1995 J. Low Temp. Phys. {\bf 101} 537,
and private communication

\bibitem{TuckLT}
Tucker M A H  and Wyatt A F G 1996,  Czech. J. Phys. {\bf 46 S1} 263


\end{thebibliography}
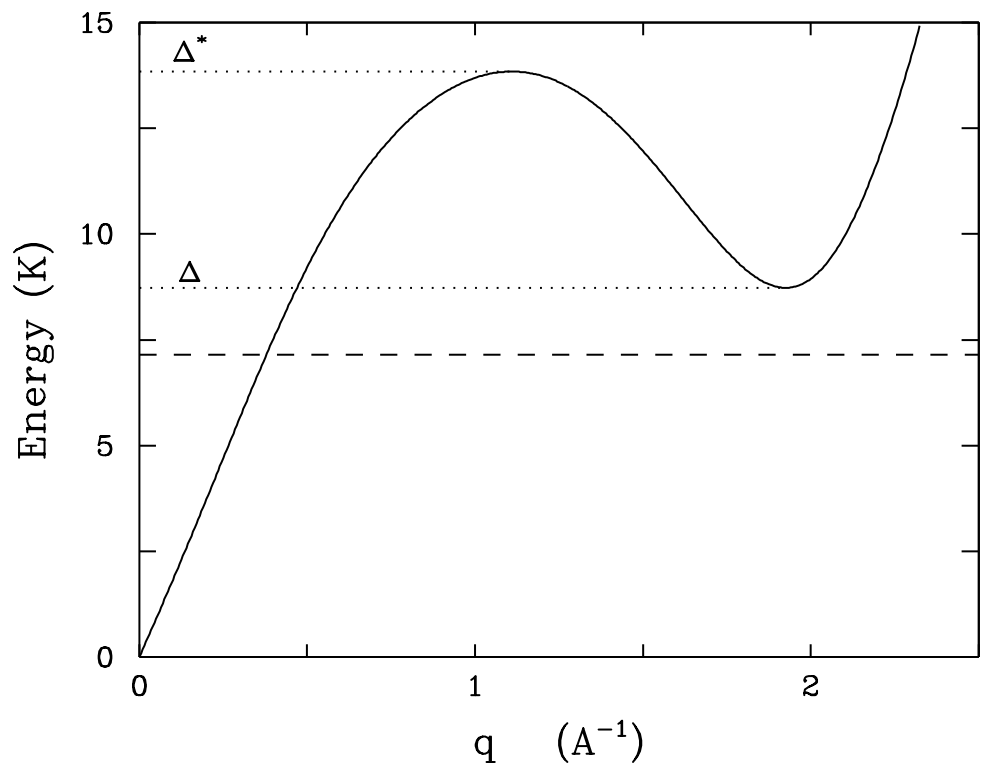
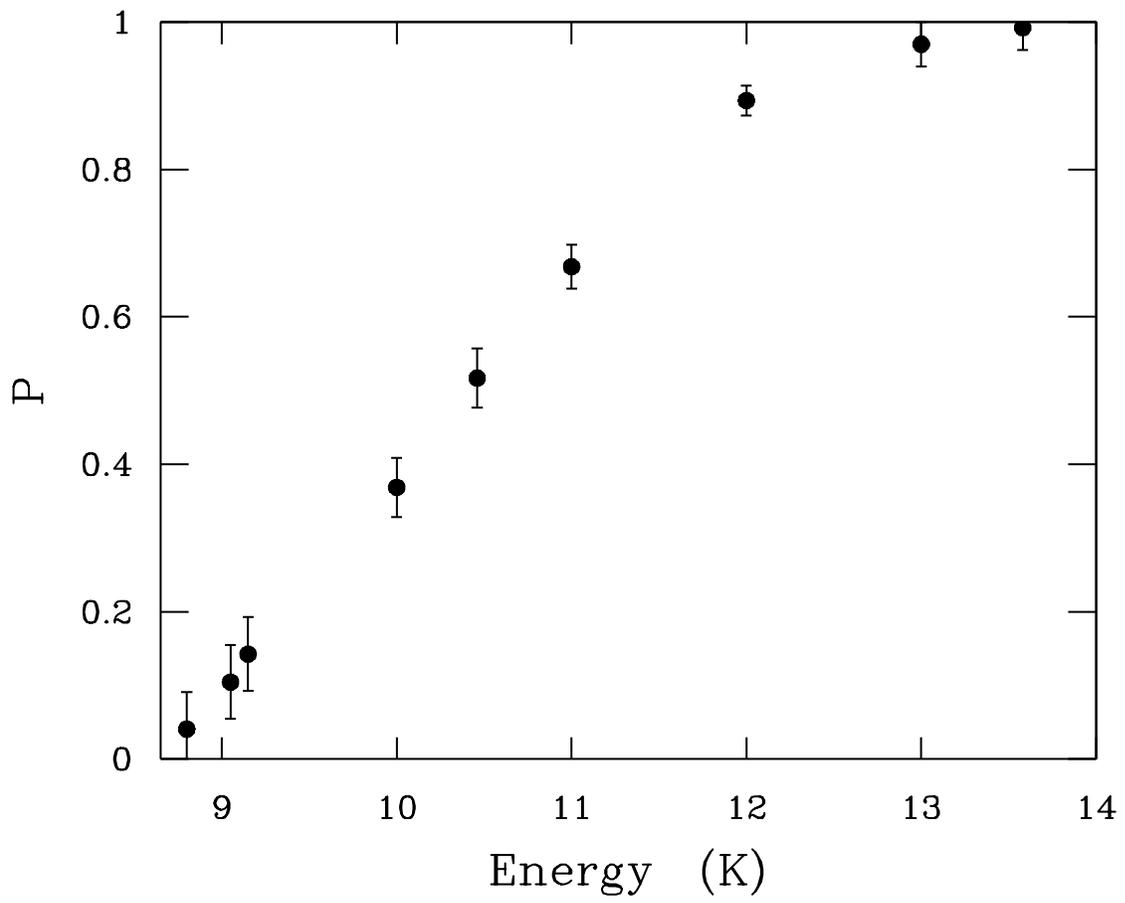
\end{document}